\documentclass[a4paper,11pt]{article}
\usepackage{pos}
\usepackage{float}
\usepackage{multirow}
\usepackage{subfigure}
\usepackage{braket}
\usepackage{tikz}

\newcommand{\beq}{\begin{equation}}
\newcommand{\eeq}{\end{equation}} 
\newcommand{\bea}{\begin{eqnarray}}
\newcommand{\eea}{\end{eqnarray}} 
\newcommand{\G}{\Gamma}
\newcommand{\csg}{SU(2)_L\times SU(2)_R}
\newcommand{\mv}{~\rm{MeV}}
\newcommand{\ggv}{~\rm{GeV}}
\newcommand{\Oop}{\mathcal{O}}

\newcommand{\q}{\overline{q}}

\newcommand{\dg}{^\dagger}
\newcommand{\gt}{\gamma_0}
\newcommand{\gx}{\gamma_1}
\newcommand{\gy}{\gamma_2}
\newcommand{\gz}{\gamma_3}
\newcommand{\gv}{\gamma_5}

\newcommand\xoh[1]{\frac{#1}{2}}
\newcommand{\oh}{\xoh{1}}

\title{Study of Chiral Symmetry and $U(1)_A$ using Spatial Correlators for $N_f=2+1$ QCD at finite temperature with Domain Wall Fermions}

\author*[a]{David Ward}
\author[b]{Sinya Aoki}
\author[c]{Yasumichi Aoki}
\author[a]{Hidenori Fukaya}
\author[d,e]{Shoji Hashimoto}
\author[c]{Issaku Kanamori}
\author[d,e,f]{Takashi Kaneko}
\author[c]{Jishnu Goswami}
\author[c]{Yu Zhang}

\affiliation[a]{Department of Physics, Osaka University,\\  Toyonaka, Osaka 560-0043, Japan}

\affiliation[b]{Center for Gravitational Physics, Yukawa Institute for Theoretical Physics, Kyoto University,\\  Kyoto 650-0047, Japan}
  
\affiliation[c]{RIKEN Center for Computational Physics(Riken CCS),\\  Kobe 650-0047, Japan}
  
\affiliation[d]{KEK Theory center, High Energy Accelerator Research Organization(KEK),\\ Tsukuba 305-0801, Japan}

\affiliation[e]{School of High Energy Accelerator Science, Graduate University for Advanced Study(SOKENDAI),\\ Tsukuba 305-0801, Japan}

\affiliation[f]{Kobayashi-Maskawa Institute for the Origin of Particles and the Universe, Nagoya University,\\ Aichi 464-8603, Japan}

\renewcommand{\logo}{\relax}
\renewcommand{\hookAfterAbstract}{%
   \par\bigskip
   \textsc{Preprint Number}:
   \href{}{KEK-CP-0398,}
   \href{https://arxiv.org/abs/2401.07514}{\color{black}{arXiv:2401.07514}}
}

\emailAdd{ward@het.phys.sci.osaka-u.ac.jp}

\abstract{Based on simulations of 2+1 flavor lattice QCD with M\"obius domain wall fermions at high temperatures, we compute a series of spatial correlation functions to study the screening masses in mesonic states. We compare these masses with the symmetry relations for various quark masses and lattice sizes at temperatures above the critical point. Using these spatial correlation functions we examine the $SU(2)_L \times SU(2)_R$ symmetry as well as the anomalously broken axial $U(1)_A$ symmetry. Additionally we explore a possible and emergent chiral-spin symmetry $SU(2)_{CS}$.}

\FullConference{The 40th International Symposium on Lattice Field Theory (Lattice 2023)\\
July 31st - August 4th, 2023\\
Fermi National Accelerator Laboratory\\}

\begin{document}
\maketitle

\section{Introduction}
\label{sec:Intro}

Chiral symmetry breaking and the axial $U(1)$ symmetry (broken by quantum anomaly) are of interest to the study of Quantum Chromodynamic (QCD) phase transition. While the role that $SU(N_f)_L\times SU(N_f)_R$ plays is more straightforward as chiral symmetry breaking occurs at lower temperatures in QCD\cite{Borsanyi:2010bp,Bazavov:2011nk,Detar:1987hib}. $U(1)_A$ is broken by anomaly below the QCD cutoff scale, and so is considered to be insensitvie to changes in both temperature as well as quark mass. However, the relationship between these two symmetries, which some work has shown to be relevant through the chiral condensate, may in fact play more of a role in how the chiral phase transition occurs, which is of particular importance in our continued interest in deconfined quark matter and QCD at extreme temperatures such as those in the early universe.\par
Previous studies by JLQCD \cite{Aoki:2021qws,Fukaya:2017wfq,Suzuki:2019vzy,Suzuki:2020rla,Rohrhofer:2019qwq}, as well as others \cite{Brandt:2016daq,Datta:2012fz}, of $N_f=2$ QCD have shown both chiral symmetry restoration, and suppression of susceptibility to the anomalously broken $U(1)_A$ symmetry, for a large range of quark masses above $T_c$ \cite{Tomiya:2016jwr}. In addition to this, it has been reported that simulations of $N_f=3$ QCD have shown similar behaviors \cite{Laudicina:2022ghk,DallaBrida:2021ddx}. Therefore, it is of interest to explore $N_f=2+1$ QCD as a way to understand more fully the roles of these symmetries in full QCD \cite{Bazavov:2019www}. While $N_f=2$ simulations have shown the aforementioned behaviors, and thus, some evidence that $\csg$ and $U(1)_A$ symmetries are connected. $U(1)_A$ appears to play a role in the chiral symmety breaking, it is unclear if this effect is more suppressed in full QCD making $N_f=2+1$ important to study.\par 
At the same time that we see this restoration in symmetries for fermions above the psuedo critical temperature, it has been pointed out that an additional set of possible symmetries emerge for very high temperatures above $T_c$. We also comment on this as these symmetries may appear in the $N_f=2+1$ simulations as we simulate higher temperatures \cite{Rohrhofer:2017wyg}.\par
In this presentation of the results of the most recent simulations of $N_f=2+1$ QCD, we begin with a short introduction of the relevant observables and the spatial correlation functions in section \ref{sec:Mesons}. Following this, in section \ref{sec:Simulations} we review our fitting methods, present our effective masses and finally the symmetry behaviors for quark masses from the physical quark masses up to $30\mv$ and temperatures in the range $136\mv-204\mv$. In addition to this we summarize our numerical results for the difference in the mesons screening masses in different channels and compare the results of the measurements for $N_f=2+1$ with those of the $N_f=2$ measurements presented in the previous work, done by JLQCD \cite{Aoki:2021lc}.\par

\section{Mesonic Correlators}
\label{sec:Mesons}
For $N_f=2+1$ the mass of the stange quark is much heavier than the up and down quarks, and breaks $SU(3)$ symmetry, making the relevant chiral symmetry $\csg$. $U(1)_A$ is known to be broken by anomaly, but it has been suggested by Pisarski and Wilczek \cite{Pisarski:1983ms} that the disappearance of this anomaly may affect the universality of the QCD phase transition; for this reason we also study the axial anomaly and its temperature dependence. Additionally, we observe that at very high temperatures an approximate symmetry emerges as a result of the large Matsubara induced mass. This corresponding $SU(2)$ symmetry  called $SU(2)$ chiral-spin or $SU(2)_{CS}$ is examined in this work in the following section.\par
\subsection{Spatial Mesonic Correlators}
\label{ssec:SpMesons}
Let us consider the quark flavor triplet bilinear operators 
\beq
\Oop_\G(x)=\q (x)(\G\otimes \xoh{\vec{\tau}})q(x) \label{eqn:Oopdef},
\eeq
where $\tau^a$ is an element of the generators of $SU(2)$, $\G$ denotes the combination of Dirac $\gamma$ listed in Table. \ref{tab:corrops}. Using the correlation functions of the $\Oop_\G(x)$ operators we examine symmeries at high temperature using the screening mass spectrum. Specifically we make use of the two point correlator along the $z$-axis,
\beq\label{eqn:zdspatialcorr}
C_\G(n_z)=\sum_{n_t,n_x,n_y} \braket{\Oop_\G(n_x,n_y,n_z,n_t)\Oop\dg_\G(0,0,0,0)}
\eeq
were $n_t$ is the Euclidean time direction and we specify the correlator channel with $\G$. At long distances the correlation function is expected to exponentially decay following $C_\G(z)\sim e^{-M_{\G}z}$, where $M_\G$ is the screening mass of the lowest energy state in the $\G$ channel.\par
As the quarks are subject to anti-periodic boundary conditions in the temporal direction, the momentum recieves a large contribution from the Matsubara frequencies. If the largest contribution to the two-point correlation function is from a pair of static non-interacting quarks, then the screening mass $M_\G$ is well approxmiated to $2\pi T$, twice the lowest Matsubara frequency. Therefore, a comparison between these two quanitites $M_\G$ and $2\pi T$ provides an effective estimation of the strength of interaction between quarks in QCD.

\begin{table}[h]

   \centering
   \begin{tabular}{c c c c c}
      \hline
      $\G$ & Reference Name & Abbr. & Associated Symmetry\\
      \hline
      $\mathbb{I}$ & Scalar & S & \multirow{2}{*}{$U(1)_A$}\\
      $\gv$ & Psuedo Scalar & PS\\
      $\gamma_k$ & Vector & $\mathbf{V_k}$ & \multirow{2}{*}{$SU(2)_L \times SU(2)_R$} & \multirow{4}{*}{$\bigg\} SU(2)_{CS}?$ }\\
      $\gamma_k\gv$ & Axial Vector & $\mathbf{A_k}$\\
      $\gamma_k\gz$ & Tensor &  $\mathbf{T_k}$ & \multirow{2}{*}{$U(1)_A$} \\
      $\gamma_k\gz\gv$ & Axial Tensor &  $\mathbf{X_k}$\\
      \hline
   \end{tabular}
   \caption{Here we express the set of observables associated with the quark correlation functions, and the symmetry group transformations which connect the various operators. In this table we have chosen to project onto states in the $z$-direction.}
   \label{tab:corrops}
\end{table}

\subsection{Emergence of $SU(2)_{CS}$}
Let us consider a quark propagator which has momenta $(p_x, p_y)$ perpendicular to the $z$-axis:
\beq
\braket{\q(z)q(0)}(p_1,p_2)=\sum_{p_0}\int_{-\infty}^\infty \frac{dp_3}{2\pi}\dfrac{m-(i\gt p_0+i\gamma_ip_i)}{p_0^2+p_i^2 +m^2}e^{ip_3z},
\eeq
where we have decomposed the Euclidean time direction into the Matsubara frequencies $p_0=2\pi T(n+\oh)$ and labelled $x,y$ and $z$ momenta with numerical indicies. We can integrate spatially by taking the $z$-directional momentum corresponding to the ``energy spectrum'' $E=\sqrt{p_0^2+m^2+p_1^2+p_2^2}$, 
\beq
\braket{\q(z)q(0)}(p_1,p_2)=\sum_{p_0}\dfrac{m+\gz E-(i\gt p_0+i\gx p_1+i\gy p_2)}{2E}e^{-Ez}.
\eeq
At a very high temperature we can take $T\gg m^2+p_1^2+p_2^2$ and expand the correlator in terms of $1/T$:
\beq
\braket{\q(z)q(0)}=\gz\xoh{1+i \mbox{sgn}(p_0)\gt\gz}e^{-\pi Tz}+O(1/T).
\eeq
Here we have approximated the summation over $p_0$ to be the lowest Matsubara mode only. This form of the correlator is invariant under an additional set of transformations:
\beq
\begin{split}
q(x)&\rightarrow e^{i\Sigma^a\theta^a}q(x)\\
\q (x)&\rightarrow e^{i\Sigma^a\theta^a}\q(x),
\end{split}
\eeq
where the set of transformations is defined by
$$\Sigma = \begin{bmatrix}
\gv\\
\gx\\
\gy
\end{bmatrix}.$$
The set of the generators of the so-called $SU(2)$ chiral-spin or $SU(2)_{CS}$ symmetry, explored in \cite{Glozman:2016swy,Glozman:2015qzf,Rohrhofer:2019saj, Rohrhofer:2019qwq,Bala:2023iqu,Chiu:2023hnm}.

\section{Numerical Results}
\label{sec:Simulations}
Our simulations of $N_f=2+1$ flavor QCD employ the tree-level improved Symanzik gauge action, as well as M\"obius domain wall fermion action with stout smearing. For all ensembles $\beta=4.17$ with a lattice cutoff set to $a^{-1}=2.453\ggv$. The residual mass of the domain-wall quark is reduced to $<1\mv$. Table \ref{tab:nf2+1params} lists our simulated ensembles and their parameters. All lattices are fixed to $L=32$, with the exception of the $36^3 \times 18$ ensembles, which keep the $L/T$ aspect ratio to greater than $2$. Our up and down quark masses begin around the physical point at $5\mv$ and take values up to $30\mv$, while the strange quark mass is fixed at around the physical value. Temperatures for the various ensembles are listed in Table \ref{tab:nf2+1params} and span the interval $T=136\mv - 204\mv$. We expect that the range of temperatures covers from $0.9T_c$ to $1.3T_c$.\par
\begin{figure*}[h]
\subfigure[~]{
\includegraphics[scale=0.47]{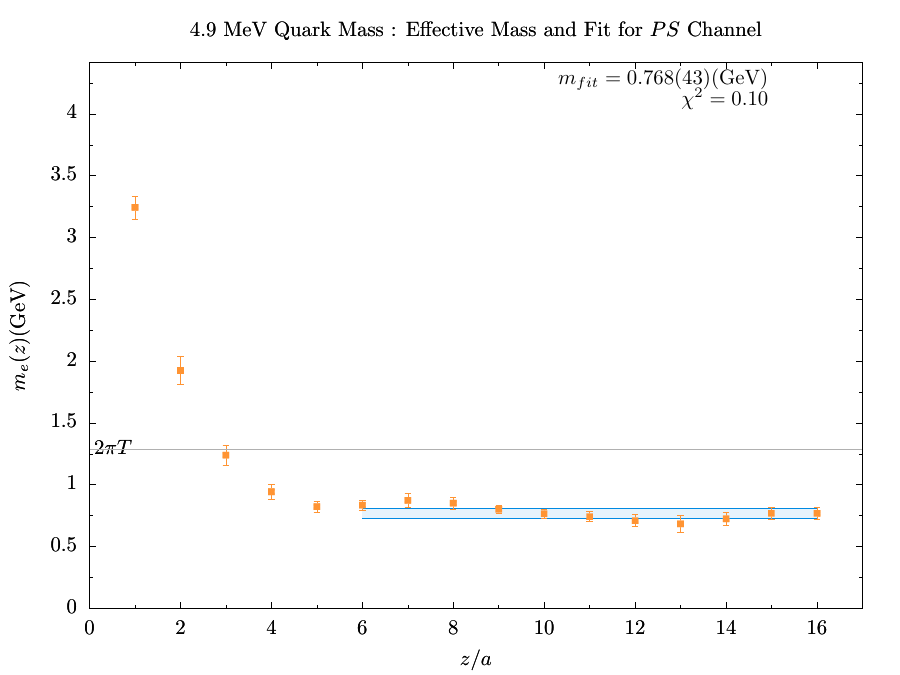}
\label{fig:204PS0020}
}
\subfigure[~]{
\includegraphics[scale=0.47]{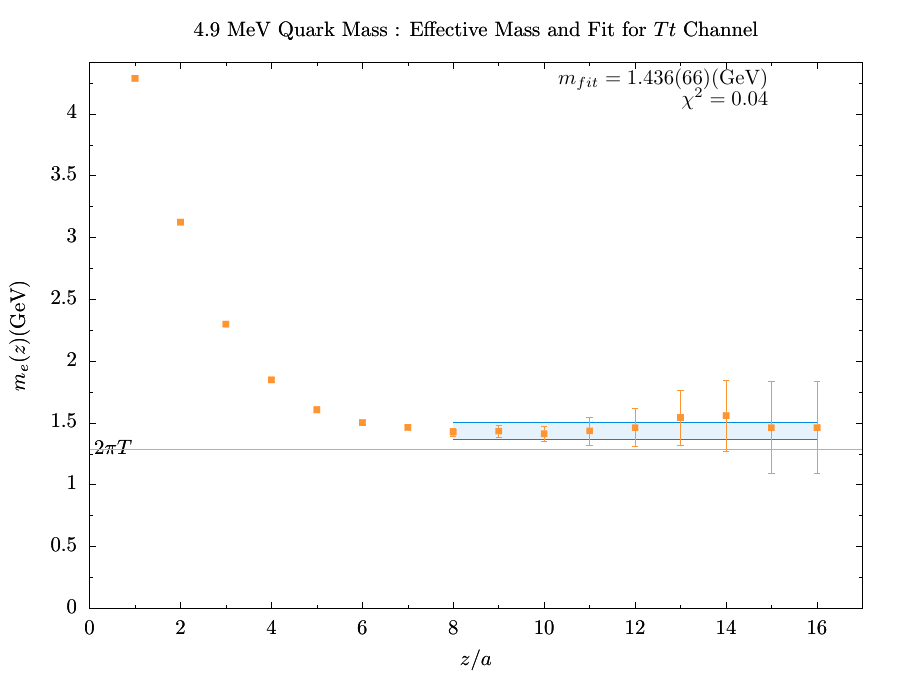}
\label{fig:204Tt0020}
}
\caption{Above are the effective mass plots for the $PS$ and $Tt$ channels with $m=0.0020$ at $T=204\mv$.
\label{fig:32x12coshfits}}
\end{figure*}

The value for the screening mass is extracted from fitting lattice data to the standard form
\beq\label{eqn:coshfits}
C(z)=A\cosh(m[z-L/2]).
\eeq
Figure \ref{fig:32x12coshfits} is an example of effective mass for the $PS$ \ref{fig:204PS0020} and $Tt$ \ref{fig:204Tt0020} channels, which are taken from our lightest mass $m=0.0020$ ensemble at temperature $T=204\mv$. Lattice points which are symmetric with respect to the reflection around the midpoint on the $z$-axis are averaged. The shaded region encompassing the fitting range shows the estimated range of the screening mass within error. A thin horizontal line is included in the plot to show the value of twice the Matsubara mass $2\pi T$.

\begin{table}[ht]
   \centering
    \begin{tabular}{c c c c c}
      $\beta$ & $T\quad(\rm{MeV})$ & $L^3\times L_t$ &  $am$ & $m\quad(\rm{MeV})$\\
      \hline\hline
      4.17 & 204 & $32^3\times 12$ & 0.0020 & 4.9\\
      & & $32^3\times 12$  & 0.0035 & 8.6\\
      & & $32^3\times 12$  & 0.0070 & 17\\
      & & $32^3\times 12$  & 0.0120 & 29\\
      \hline
      & 175 & $32^3\times 14$  & 0.0020 & 4.9\\
      & & $32^3\times 14$  & 0.0035 & 8.6\\
      & & $32^3\times 14$  & 0.0070 & 17\\
      & & $32^3\times 14$  & 0.0120 & 29\\
      \hline
      & 153 & $32^3\times 16$ & 0.0020 & 4.9\\
      & & $32^3\times 16$  & 0.0035 & 8.6\\
      & & $32^3\times 16$  & 0.0070 & 17\\
      & & $32^3\times 16$  & 0.0120 & 29\\
      \hline
      &136 & $36^3\times 18$ & 0.0020 & 4.9\\
      & & $36^3\times 18$  & 0.0035 & 8.6\\
      & & $36^3\times 18$  & 0.0070 & 17\\
      & & $36^3\times 18$  & 0.0120 & 29\\
      \hline
   \end{tabular}
   \caption{Parameters of the measured lattices for $N_f=2+1$ QCD for the various mass ensembles and temperatures are given above. The physical quark mass is fixed to the physical value, the psuedocritical transition temperature is $T_c\sim 155\mv$.}
   \label{tab:nf2+1params}
\end{table}

From the fitted mass values determined in Figure \ref{fig:32x12coshfits}, we evaluate symmetry breaking by the screening mass difference between the partner channels related by the transformations $\Gamma$ listed in Table \ref{tab:corrops}. 
For a mass difference of $\sim 0.5-1.0\ggv$, which is on the scale of the screening mass itself, the symmetry is considered strongly broken.\par
\begin{figure}[h!]
\centering
\scalebox{0.7}{
\begin{tikzpicture}
\node (A) {\subfigure[~]{\includegraphics[width=\textwidth]{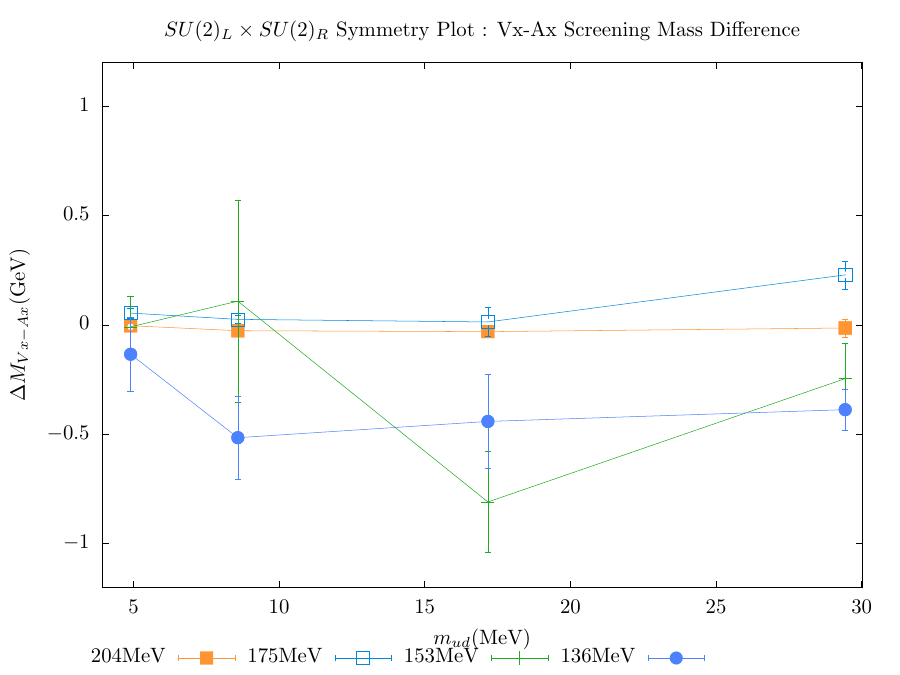}\label{sfig:compVx-Ax2+1}}};
\node[anchor=north east, yshift=-0.95cm, xshift=-0.85cm] at (A.north east)  {\subfigure[~]{\includegraphics[width=.4\textwidth]{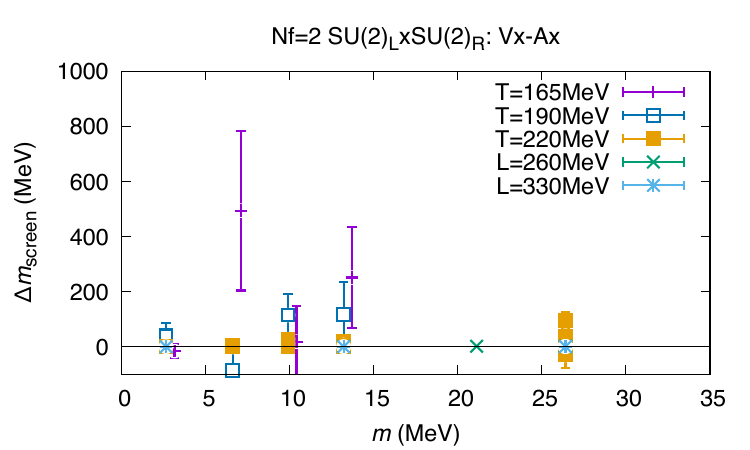}\label{sfig:compVx-Ax2}}};
\end{tikzpicture}
}
\caption{\ref{sfig:compVx-Ax2+1} is a plot of the $SU(2)$ chiral symmetry which is intact at all temperatures in the chiral limit; however, as the quarks become increasingly massive we have broken symmetries emerge for $153\mv$ and $136\mv$ ensembles while the higher temperature $204\mv$ is still in a largely unbroken state. This appears to indicate a restoration of chiral symmetry for temperatures well above the psuedocritical temperature.\\
\label{fig:SU2nf2vnf2+1}}
\end{figure}
\begin{figure*}[hb!]
\centering
\scalebox{0.7}{
\begin{tikzpicture}
\node (A) {\subfigure[~]{\includegraphics[width=\textwidth]{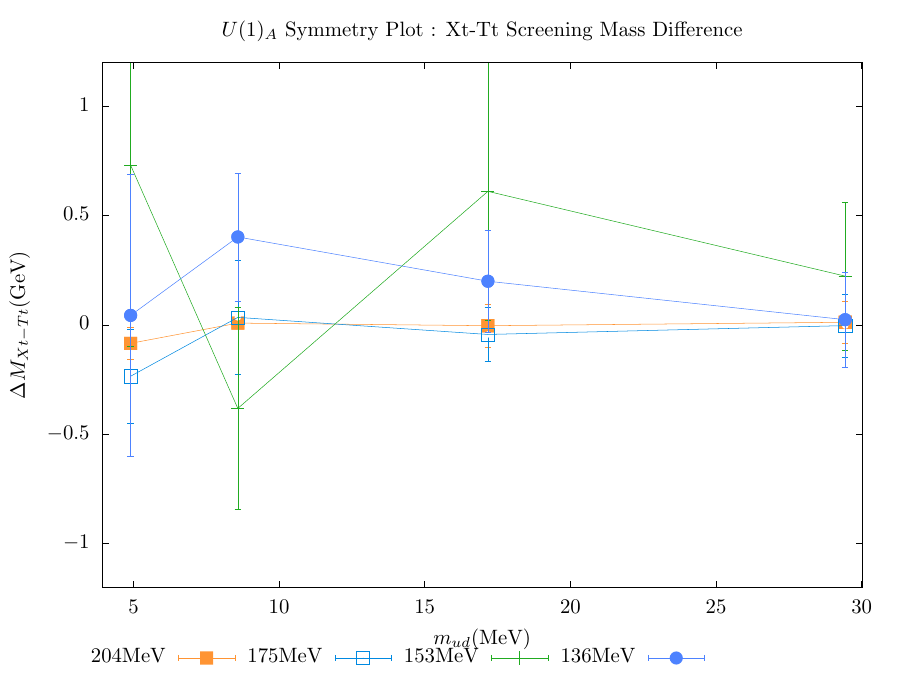}\label{ssfig:compXt-Ttnf2+1}}};
\node[anchor=south east, yshift=2.1cm, xshift=-0.9cm] at (A.south east)  {\subfigure[~]{\includegraphics[width=.4\textwidth]{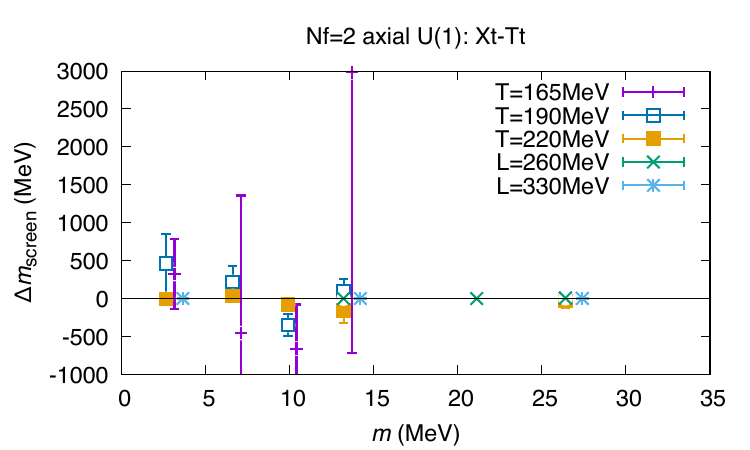}\label{sfig:compXt-Ttnf2}}};
\end{tikzpicture}
}
\caption{As stated in table \ref{tab:corrops}, figure \ref{ssfig:compXt-Ttnf2+1} associated with the the $U_A(1)$ symmetry of the lattice. The behavior shown by the high mass quarks and high temperature ensembles is consistent with the $U_A(1)$ susceptibility insensitvity. 
\label{fig:U1nf2vnf2+1}}
\end{figure*}
\begin{figure}[t]
\centering
\includegraphics[scale=0.7]{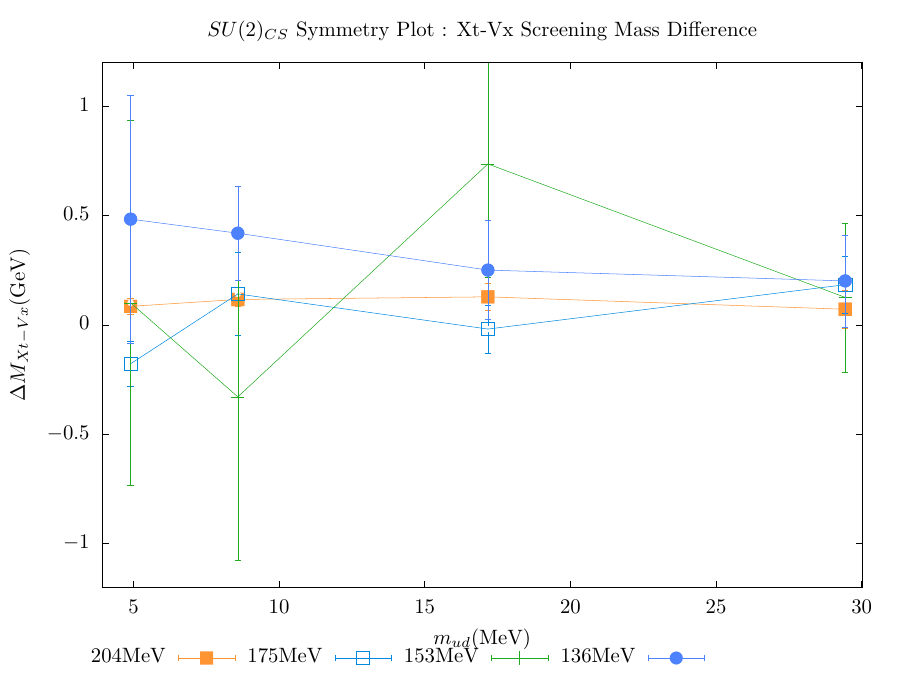}
\caption{Above is the creening mass difference plot for $SU(2)_{CS}$ for the temperature ensembles and mass ranges listed in Table \ref{tab:nf2+1params}. \label{fig:Xt-Vxnf2+1sym}}
\end{figure}

Figures \ref{fig:SU2nf2vnf2+1}, \ref{fig:U1nf2vnf2+1}, and \ref{fig:Xt-Vxnf2+1sym} show the mass differences for the symmetries listed in Table \ref{tab:corrops}. Because the mass difference can take on a negative value without affecting the overall symmetry behavior, as seen in the $N_f=2$ plot of figures \ref{sfig:compVx-Ax2} and \ref{sfig:compXt-Ttnf2}, the simple mass difference of the $N_f=2+1$ is plotted.\par
For the $N_f=2+1$ plot of $\csg$ in Figure \ref{fig:SU2nf2vnf2+1}, the ensembles at $T=204\mv$ and $175\mv$ have a mass difference consistent with zero for a wide range of $m_{ud}$. In particular the largest difference at $m=0.007$ for $T=204\mv$ is only $\approx 33\mv$. Likewise for $175\mv$, with the exception of the highest mass, which has not undergone transition, all values are within $30\mv$. For lower temperatures, the screening mass differences for higher mass ensembles increase and are on the order of $\sim 0.5-1 \ggv$, however, all of the lightest mass quarks are consistent with zero to within $O(100\mv)$ error bars.\par

For comparison we also plot in Fig. \ref{sfig:compVx-Ax2} our previous result for $N_f=2$ QCD \cite{Rohrhofer:2019qwq}. Taking into account $T_c\sim 165\mv$ in $N_f=2$ QCD\footnote{See JLQCD Collaboration, in proceedings by H.Fukaya.} the plot shows a similar behavior to our $N_f=2+1$ results.
In comparions with the plot for $\csg$ the plot of $U(1)_A$ Fig. \ref{fig:U1nf2vnf2+1} has larger errors, but appears to show a similar tendency. For the highest temperature ensemble $T=204\mv$, the $Xt-Tt$ mass difference is consistent with zero for entire range of masses. A behavior which $T=175\mv$ appears to also mirror, but with increased noise. For lower temperatures we again see large mass differences on the order of $\sim 0.5-1 \ggv$.
Compared to the $N_f=2$ results in \ref{sfig:compXt-Ttnf2} the behaviors do overlap, with temperatures above $1.1T_c$ showing strong violation of the symmetry.\par
The plot of the emergent $SU(2)_{CS}$ symmetry in \ref{fig:Xt-Vxnf2+1sym} appears to be broken for the $153\mv$ and $136\mv$ ensembles while $204\mv$ again appears to be in a state where a symmetry is accessible. However, due to the increased noise in this plot it is less clear if there is a symmetry present, as all points except $T=204\mv$ at the physical point are too noisy to draw conclusions about their overall behavior.\par

\section{Summary of Preliminary Results}
\label{sec:Conclusion}
Based on the evidence of symmetry restorations in plots \ref{sfig:compVx-Ax2+1}, and \ref{ssfig:compXt-Ttnf2+1} we can conclude that for temperatures above $1.1T_c$ there is likely a restoration of chiral symmetry as well as a suppression of the axial $U(1)$ broken symmetry. This is consistent with the previous evidence in the $N_f=2$ measurements. As noise is a key issue for both sets of measurements further refinement of measurements for $N_f=2+1$ will help clarify the behavior of $U(1)_A$, as well as solidify our initial conclusions about $\csg$ symmetry restoration.\par
Unfortunately the current outlook on $SU(2)_{CS}$ is unclear as the measurements presented in the previous section appear to also suffer from noise, and thus if there is a symmetry present it is harder to detect. From the observation in section \ref{ssec:SpMesons}, it is possible the symmetry emerges within our range of measurement, but it is likely to occur at very high temperatures.\par
As for the current measurements, we may improve the precision and strengthen our results by increasing statistics. One possible path forward is to use rotational invariance of the $z$-directional states to take additional measuremnts of the $x$ and $y$ polarizations of the correlation function thus giving us a larger set of statistics to reduce the long range noise of the effective mass plots and fits. We also plan to include low mode averaging, where the source points of the low lying mode's contribution to the Dirac operators is averaged.\pagebreak
\acknowledgments{
We thank L. Glozman and Y. Sumino for useful discussions. For the numerical simulation
we use the QCD software packages Grid \cite{Boyle:2015tjk,Meyer:2021uoj} for configuration generations and Bridge++ \cite{Ueda:2014rya,Akahoshi:2021gvk} for measurements. Numerical simulations were performed on Wisteria/BDEC-01 at JCAHPC under a support of the HPCI System Research Projects (Project ID: hp170061) and Fugaku computer provided by the RIKEN Center for Computational Science under a support of the HPCI System Research Projects (Project ID: hp210231). This work is supported in part by the Japanese Grant-in-Aid for Scientific Research (No. JP26247043, JP18H01216, JP18H04484, JP18H05236,JP22H01219) and by Joint Institute for Computational Fundamental Science (JICFuS).}

\bibliographystyle{/usr/share/texlive/texmf-dist/bibtex/bst/base/JHEP}
\bibliography{skele-sym}

\end{document}